\documentclass[twocolumn,pra,superscriptaddress,showpacs]{revtex4}
\usepackage{amssymb}
\usepackage{amsmath}
\usepackage{graphicx}
\usepackage{bm}

\begin{document}
\title{Modal, Spectral, and Polarization Entanglement in Guided-Wave Parametric Down-Conversion}

\author{Mohammed F. Saleh}
\email{mohsaleh@bu.edu}\affiliation{Quantum Photonics Laboratory,
Department of Electrical and Computer Engineering, Boston
University, Boston, MA 02215}

\author{Bahaa E. A. Saleh}
\affiliation{Quantum Photonics Laboratory, Department of Electrical
and Computer Engineering, Boston University, Boston, MA 02215}
\affiliation{Quantum Photonics Laboratory, College of Optics and
Photonics (CREOL), University of Central Florida, Orlando, FL 32816}

\author{Malvin Carl Teich}
\affiliation{Quantum Photonics Laboratory, Department of Electrical
and Computer Engineering, Boston University, Boston, MA 02215}
\affiliation{Department of Physics, Boston University, Boston, MA
02215}
\date{\today}

\begin{abstract}
We examine the modal, spectral, and polarization entanglement
properties of photon pairs generated in a nonlinear, periodically
poled, two-mode waveguide (1-D planar or 2-D circular) via
nondegenerate spontaneous parametric down-conversion.  Any of the
possible degrees of freedom --- mode number, frequency, or
polarization --- can be used to distinguish the down-converted
photons while the others serve as attributes of entanglement.
Distinguishing the down-converted photons based on their mode
numbers enables us to efficiently generate spectral or polarization
entanglement that is either narrowband or broadband. On the other
hand, when the generated photons are distinguished by their
frequencies in a Type-0 process, modal entanglement turns out to be
an efficient alternative to polarization entanglement. Moreover,
modal entanglement in Type-II down-conversion may be used to
generate a doubly entangled state in frequency and polarization.
\end{abstract}
\pacs{42.65.Wi, 42.65.Lm, 42.50.Dv}
\maketitle

\pagestyle{empty}
\section{Introduction}
Photon pairs generated via the nonlinear process of spontaneous
parametric down-conversion (SPDC) can exhibit entanglement in
multiple degrees of freedom: spectral, spatial, and polarization
\cite{klyshko80}. Entanglement arises from the multiple
possibilities for satisfying energy and momentum conservation, as
required by the parametric interaction process. Spectral entanglement 
is exhibited equivalently in time or energy \cite{saleh00}; 
spatial entanglement may be manifested as entanglement in wavevector
direction \cite{joobeur94}, 
transverse direction \cite{joobeur96}, 
orbital-angular-momentum (OAM) \cite{mair01}, or spatial parity
\cite{abouraddy07,yarnall07a}. A proper description of the quantum
state of the photon pair (biphoton) requires a Hilbert space that
includes all of the pertinent degrees of freedom. One of these
degrees of freedom may be arbitrarily chosen to distinguish the two
photons, while the other two are used to describe the state. The
term hyper-entanglement has been used to describe states defined
over multiple degrees of freedom \cite{hyperentangled}.

While the polarization variable is intrinsically binary, spectral
and spatial variables are generally continuous. Nevertheless, binary
spectral or spatial degrees of freedom can be extracted by selection
of a subspace of dimension two (a qubit). Examples are frequency
selection by use of narrow spectral filters
\cite{filterspectralentangl}, direction selection by use of pinholes
in the far field \cite{directionalentangl}, and lower-order OAM-mode
selection by suppression of higher-order modes
\cite{transverseentangl}. Another approach to binarization is based
on constraining the parametric process in such a way that only two
values of the continuous variable are permitted; such configurations
include narrowband spectral entanglement \cite{counterSPDC} and
Type-II noncollinear down-conversion in which only two directions
(as determined by intersecting rings) are permitted
\cite{polentanglement}. A third approach is to map the larger
Hilbert space onto a binary space, as in spatial-parity entanglement
\cite{abouraddy07,yarnall07b}.

In this paper, we consider the generation of biphotons by means of
SPDC in a nonlinear waveguide. This configuration confines the
photons to a single direction of propagation and discretizes the
spatial degree of freedom to the waveguide spatial modes.  For a
two-mode waveguide, the spatial degree of freedom is binary,
representing a modal qubit. The waveguide also supports two
polarizations (e.g., TE and TM in a one-dimensional geometry).
Moreover, the spectral distributions of the down-converted photons
may be constrained to yield a pair of narrow spectral lines defining
a spectral qubit, or alternatively, may be made more flexible so as
to generate photon pairs with broad spectral entanglement; these
conditions are achieved by making use of periodic and linearly
chirped poling of the nonlinear medium, respectively.

In addition to the advantage of spatial binarization, the two-mode
waveguide has the merit of combining a higher rate of biphoton
generation, normally obtained in collinear degenerate bulk SPDC,
with the photon separability generally offered by a noncollinear
configuration. Moreover, processing of the generated biphotons for
applications in quantum information \cite{Qinformation} is often
facilitated by the use of guided-wave devices.

In this paper, we offer a comprehensive theoretical study of the
properties and applications of modal, spectral, and polarization
entanglement of biphotons generated via SPDC in 1-D planar and 2-D
circular waveguides. In particular, we consider biphotons generated
via different types of interactions and with the use of a continuous
wave (CW) pump. Prior work in this area has been limited and has
made use of a pulsed, rather than CW, pump
\cite{controlledspatialmodes,modeentanglement}.

The paper is organized as follows. In Sec.~II, we develop a general
theory for SPDC in periodically poled multimode waveguides. In
Sec.~III, we determine the waveguide parameters required to generate
modal and spectral or polarization entanglement in 1-D planar
waveguides using either Type-0 or Type-II interactions. A similar
study is carried out in Sec.~IV for 2-D circular waveguides (optical
fibers). Features and applications of modal entanglement
are considered in Sec.~V. Finally, the conclusions are presented in
Sec.~VI.

\section{Spontaneous parametric down-conversion in multimode waveguides}
Consider an SPDC process in a multimode waveguide, where a pump wave
$\,p\,$  is down-converted into a signal wave $\,s\,$  and an idler
wave $\,i\,$. Using time-dependent perturbation theory, the
two-photon state $\vert \Psi\rangle$ can be written as
\cite{counterSPDC}
\begin{equation}
\begin{array}{l}
\vert \Psi\rangle\thicksim \displaystyle \iint
\mathrm{d}\mathbf{r}\, \mathrm{d}t\;\mathbf{d}(\mathbf{r})\,
{\mathbf{E}}_{p}(\mathbf{r},t)\,
\widehat{\mathbf{E}}^{(-)}_{s}(\mathbf{r},t)\,
\widehat{\mathbf{E}}^{(-)}_{i}(\mathbf{r},t) \,\vert 0,0\rangle,
\end{array}
\label{eq3}
\end{equation}
where $\mathbf{d}(\mathbf{r})$ is the second-order
nonlinear-coefficient tensor; $\mathbf{E}_p$ is the pump electric
field at position $\mathbf{r}$ and time $t$, treated classically and
assumed to be undepleted; $\widehat{\mathbf{E}}^{(-)}_{q}$ are the negative-frequency parts of the signal and idler
electric-field operators ($q= s,\,i$) at position $\mathbf{r}$ and
time $t$; and $\vert 0,0\rangle$ is the vacuum state of the signal
and the idler.

For guided waves propagating along the $x$ direction, the
electric-field operator of the signal and idler is written as
\begin{equation}
\begin{array}{cl}
\widehat{\mathbf{E}}^{(-)}_{q}(\mathbf{r},t)= \displaystyle\int
\mathrm{d}\omega_{q}\!\sum_{m_{q},\sigma_{q}}&
\!\!\widehat{a}^{\dagger}_{m_{q},\sigma_{q}}\!\left(\omega_{q}\right) \,u_{m_{q},\sigma_{q}}\left( \omega_{q},y ,z\right) \\
&\times\exp \!\left[j\omega_{q}t- j\beta_{m_{q},\sigma_{q}}(\omega_{q})\,x 
\right], \label{eq4}
\end{array}
\end{equation}
where $\omega_{q}$ is the angular frequency; $\sigma_{q}$ and
$m_{q}$ are the polarization and spatial-mode indexes, respectively;
$\beta_{m_{q},\sigma_{q}}$ is the propagation constant;
$u_{m_{q},\sigma_{q}}$ is the transverse field profile in the
$y$-$z$ plane; and $\widehat{a}_{m_{q},\sigma_{q}}^{\dagger}$ is the
creation operator for the wave $q$. For the classical pump electric field, a similar expression can be obtained by taking the complex conjugate of Eq.~(\ref{eq4}) then replacing the creation operator with the wave amplitude.

Substituting Eq.~(\ref{eq4}) into Eq.~(\ref{eq3}), assuming that
$\mathbf{d}(\mathbf{r})$ depends only on $x$, and noting that the
integration over $t$ yields the delta function $\delta\left(
\omega_{p}-\omega_{s} -\omega_{i} \right)$, the two-photon state
becomes
\begin{equation}
\begin{array}{c}
\vert \Psi\rangle\thicksim \displaystyle \int
\mathrm{d}\omega_{s}\:\sum_{\mathbf{m},\boldsymbol{\sigma}}
\:\Phi_{\mathbf{m},\boldsymbol{\sigma}}\left(
\omega_{s}\right)\:\vert
\omega_{s},m_{s},\sigma_{s}\rangle\vert\omega_{i},m_{i},\sigma_{i}\rangle,
\end{array}
\label{eq5}
\end{equation}
where
\begin{equation}
\begin{array}{c}
\Phi_{\mathbf{m},\boldsymbol{\sigma}}\left( \omega_{s}\right)=
A_{\mathbf{m},\boldsymbol{\sigma}}\left(
\omega_{s}\right)\,\displaystyle\int \mathrm{d}x \; \mathbf{d}(x)
\exp\left[ j\Delta\beta_{\mathbf{m},\boldsymbol{\sigma}} \left(
\omega_{s}\right)\,x\right],
\end{array}
\label{eq7}
\end{equation}
\begin{eqnarray}
A_{\mathbf{m},\boldsymbol{\sigma}}\left( \omega_{s}\right)=
\displaystyle\iint \! \mathrm{d}y\; \mathrm{d}z\;
\!\displaystyle\prod_{q=\,p,s,i}\!\!
u_{m_{q},\sigma_{q}}\left(\omega_{q}\,,y\,,z\,\right), \label{eq6}
\end{eqnarray}
$\omega_{i}=\omega_{p}-\omega_{s}$, $\mathbf{m}=(m_{s},m_{i})$,
$\boldsymbol{\sigma}=(\sigma_{s},\sigma_{i})$, and
$\Delta\beta_{\mathbf{m},\boldsymbol{\sigma}}\left(
\omega_{s}\right)=\beta_{m_{p},\sigma_{p}}\left(\omega_{p}\right)-
\beta_{m_{s},\sigma_{s}}\left(\omega_{s}\right)
-\beta_{m_{i},\sigma_{i}}\left(\omega_{i}\right) $ is the phase
mismatch. The factor $ A_{\mathbf{m},\boldsymbol{\sigma}} $
represents the spatial overlap of the transverse profiles of the
interacting modes. The square-magnitude of
$\Phi_{\mathbf{m},\boldsymbol{\sigma}}\left(\omega_{s}\right)$
represents the SPDC spectrum when the signal and the idler mode
numbers  are $\mathbf{m}=(m_{s},m_{i})$ and their polarization
indexes are $\boldsymbol{\sigma}=(\sigma_{s},\sigma_{i})$.

Since the generated photons are collinear, their directions cannot
be used to distinguish the two photons. If the mode number is used
in this capacity, then the two-photon state described in
Eq.~(\ref{eq5}) may be written as
\begin{equation}
\begin{array}{c}
\vert \Psi\rangle\thicksim\displaystyle \int
\mathrm{d}\omega_{s}\:\sum_{\mathbf{m},\boldsymbol{\sigma}} \:
\Phi_{\mathbf{m},\boldsymbol{\sigma}}\left( \omega_{s}\right)\:
\vert
\omega_{s},\sigma_{s}\rangle_{m_{s}}\,\vert\omega_{i},\sigma_{i}\rangle_{m_{i}}\,.
\end{array}
\label{eq51}
\end{equation}
Alternatively, using frequency as the identifier, the two-photon
state takes the form
\begin{equation}
\begin{array}{c}
\vert \Psi\rangle\thicksim\displaystyle \int\mathrm{d}\omega_{s}\:
\sum_{\mathbf{m},\boldsymbol{\sigma}}
\:\Phi_{\mathbf{m},\boldsymbol{\sigma}}\left( \omega_{s}\right)\:
\vert m_{s},\sigma_{s}\rangle_{\omega_{s}}\,\vert
m_{i},\sigma_{i}\rangle_{\omega_{i}}\,.
\end{array}
\end{equation}
We may also use polarization as the identifier and write a similar
expression for the two-photon state. These expressions for the state
are, of course, equivalent.

We now derive expressions for the spectral function
$\Phi_{\mathbf{m},\boldsymbol{\sigma}}$ for quasi-phase matched
(QPM) structures. Such structures are designed to phase match a
specific type of interaction (Type 0 or I or II) via  a certain
component of  the second-order nonlinear tensor; hence $\mathbf{d}$
can be replaced by its effective value $d_{\rm eff}$. Using electric
poling, the nonlinear coefficient is made to alternate between
$\pm\, d_{\rm eff}$ along one of the crystal principle axes, say
$x$. The poling period can be either uniform or variable.  We
consider these two cases in turn.

\vspace{5pt} \noindent \textbf{Uniform Poling.} For uniform poling
with period $ \Lambda_{\,0}$,  the spatial distribution of the
nonlinear coefficient can be represented as a sum of distinct
Fourier components, $ d_{\rm eff}\left(
x\right)=\sum_{\kappa=1}^{\infty} \widetilde{d}{_{\kappa}}\,
\exp\left[-j\,\int_{0}^{x} dx\: K_{\kappa}\,(x) \right] $, where
$\widetilde{d}{_{\kappa}}=(2/\pi\kappa) \left|d_{\rm eff}\right|$
and $K_{\kappa}\,(x)= (2\pi\kappa/\Lambda_0)$ represent the
amplitude and spatial frequency of the $\kappa$th Fourier component,
respectively. Only that Fourier component whose phase is close to,
or equal to, the phase mismatch
$\Delta\beta_{\mathbf{m},\boldsymbol{\sigma}}\left(
\omega_{s}\right)$ will contribute to
$\Phi_{\mathbf{m},\boldsymbol{\sigma}}\left( \omega_{s}\right)$.
Carrying out the integration in Eq.~(\ref{eq7}) for a waveguide of
length $L$, we obtain
\begin{equation}
\begin{array}{cl}
  \Phi_{\mathbf{m},\boldsymbol{\sigma}}\left( \omega_{s}\right)=
  &A^{\prime}_{\mathbf{m},\boldsymbol{\sigma}}\left( \omega_{s}\right)\:
  \mathrm{sinc}\left[\Delta \widetilde{\beta}_{\mathbf{m},\boldsymbol{\sigma}}\left( \omega_{s}\right)L/2\pi \right]  \\
 & \times\:\exp\left[ j\,\Delta \widetilde{\beta}_{\mathbf{m},\boldsymbol{\sigma}}\left(\omega_{s}\right)L/2\right],
\end{array}
\label{eq9}
\end{equation}
where
\begin{equation}
\Delta \widetilde{\beta}_{\mathbf{m},\boldsymbol{\sigma}}\left(
\omega_{s}\right)=
\Delta\beta_{\mathbf{m},\boldsymbol{\sigma}}\left(
\omega_{s}\right)-\frac{2\pi\,
\kappa}{\Lambda_{\,0}} \,,
\end{equation}
$ A^{\prime}_{\mathbf{m},\boldsymbol{\sigma}}\left(
\omega_{s}\right)=
(2L/\pi\kappa)\left|d_{\rm
eff}\right|A_{\mathbf{m},\boldsymbol{\sigma}}\left(
\omega_{s}\right)$, and
$\mathrm{sinc}\,(\theta) =\sin(\pi\theta)/(\pi\theta)$. From a
practical perspective, the poling period is determined by satisfying
the quasi-phase-matching condition at a certain frequency
$\omega_{s} $, i.e., $\Delta
\widetilde{\beta}_{\mathbf{m},\boldsymbol{\sigma}}\left(\omega_{s}\right)=0$,
or
$\Delta\beta_{\mathbf{m},\boldsymbol{\sigma}}\left(
\omega_{s}\right)=2\pi\kappa/\Lambda_0$.

\vspace{5pt} \noindent \textbf{Linearly Chirped Poling.} For
linearly chirped poling with a slowly varying spatial frequency, we
can still make use of a Fourier series to represent the spatial
distribution of the nonlinear coefficient with $K_{\kappa}\,(x\,) =
2\pi\kappa/\Lambda(x)=  2\pi\kappa/\Lambda_0 -\zeta x$, where
$\zeta$ is the chirp parameter \cite{transformlimSH}. It is clear
that it is the spatial frequency, rather than the spatial period,
that is chirped. In this case, Eq.~(\ref{eq7}) yields
\begin{equation}
\begin{array}{cl}
\Phi_{\mathbf{m},\boldsymbol{\sigma}}\left(\omega_{s}\right)=&
A^{\prime\prime}_{\mathbf{m},\boldsymbol{\sigma}}\left(
\omega_{s}\right)\; \exp\left[ \dfrac{-j\,\Delta
\widetilde{\beta}_{\mathbf{m},\boldsymbol{\sigma}}^{2}
\left(\omega_{s}\right)}{2\zeta}\right]\vspace{1mm} \\
 & \times \: \left[\gamma\left( L\right)-\gamma\left(0\right)\right] ,
\end{array}
\label{eq8}
\end{equation}
where $A^{\prime\prime}_{\mathbf{m},\boldsymbol{\sigma}}\left(
\omega_{s}\right)=
\left|d_{\rm eff}\right|
A_{\mathbf{m},\boldsymbol{\sigma}}\left(
\omega_{s}\right)\sqrt{{2j}/{\pi\zeta
\kappa^{2}}}$, $\gamma(x)= \mathrm{erfi}\{[{\Delta
\widetilde{\beta}_{\mathbf{m},\boldsymbol{\sigma}}(
\omega_{s})+\zeta x}]/{\sqrt{-2j\zeta}}\}$,
$\mathrm{erfi}(\theta)=-j \,\mathrm{erf}(j\,\theta) $ and
$\mathrm{erf}(\cdot)$ is the error function. In the remainder of
this paper, we use $\kappa=1$ to compute the poling periods since
this results in the strongest nonlinear interaction and poling
periods are not limited by fabrication techniques.

\section{Modal entanglement in 1-D planar waveguides}
\begin{figure}
\centering
\includegraphics[width=3.4in,totalheight =1.5 in]{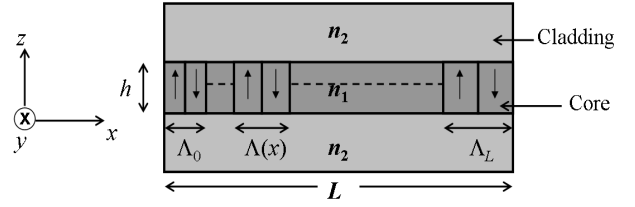}
\caption{Sketch of a 1-D planar waveguide. The waveguide is a slab
of dielectric medium of thickness $h$ with a uniform refractive
index $n_{1}$ surrounded by media of lower refractive index $n_{2}$.
The inner medium and the outer media are known as the \emph{core}
and \emph{cladding}, respectively. The nonlinear coefficient of the
core is poled with a period $\Lambda(x)$; arrows indicate the poling
direction. The poling periods at $x=0$ and $x=L$ are denoted
$\Lambda_{0}$ and $\Lambda_{L}$, respectively.} \label{figsketch}
\end{figure}

We now consider the generation of entangled photons via
nondegenerate SPDC in a 1-D planar waveguide. The structure
comprises a dielectric slab of thickness $h$ and refractive index
$n_{1}$ embedded in dielectric media with lower refractive index $
n_{2} $, as illustrated in Fig.~\ref{figsketch}. The  transverse
profile of mode $ m_{q}$ inside the slab is \cite{salehteichbook}
\begin{equation}
u_{m_{q},\sigma_{q}}\left( \omega_{q},\,y\, ,z\,\right)\varpropto\left\lbrace
\begin{array}{l}
 \cos\left( k_{m_{q},\sigma_{q}}^{(z\,)}\!\left(\omega_{q}\right)\,z\right),\quad m_{q}= {\rm even} \vspace{1.5mm} \\
 \sin\left( k_{m_{q},\sigma_{q}}^{(z\,)}\!\left(\omega_{q}\right)\,z\right),\quad\, m_{q}= {\rm odd,}
\end{array} \right.
\label{eq10}
\end{equation}
while outside the slab it is
\begin{equation}
u_{m_{q},\sigma_{q}}\left(\omega_{q},\,y\, ,z\,\right)\varpropto
\exp\left( -\gamma_{m_{q},\sigma_{q}}\!\left(
\omega_{q}\right)\,z\right),
\end{equation}
where $ k_{m_{q},\sigma_{q}}^{(z\,)}=
(k_{m_{q},\sigma_{q}}^{2}-\beta_{m_{q},\sigma_{q}}^{2})^{1/2}$ is
the \textit{z}-component of the wavevector $k_{m_{q},\sigma_{q}} =
n_{1}\,\omega_{q}/c$,
$\gamma_{m_{q},\sigma_{q}}=(\beta_{m_{q},\sigma_{q}}^{2}-\widetilde{k}_{m_{q},\sigma_{q}}^{2})^{1/2}$,
$\widetilde{k}_{m_{q},\sigma_{q}}= n_{2}\,\omega_{q}/c$, and $c$ is
the velocity of light in free space.  Note that both $ n_{1} $ and $
n_{2} $ are frequency-dependent.

Within the slab, the mode transverse profiles are either even or odd
functions. The propagation constants of the modes can be determined
using the dispersion relation for the planar waveguide. For the TE
wave ($o$-polarization),
\begin{equation}
\tan^{2}\left[ \dfrac{h}{2}\sqrt{k_{m_{q},\sigma_{q}}^{2}
-\beta_{m_{q},\sigma_{q}}^{2}} -\dfrac{m_{q}\pi}{2}\right]
=\dfrac{\beta_{m_{q},\sigma_{q}}^{2}-k_{m_{q},
\sigma_{q}}^{2}}{\widetilde{k}_{m_{q},\sigma_{q}}^{2}-\beta_{m_{q},\,\sigma_{q}}^{2}}\,,
\label{eq11}
\end{equation}
whereas for the TM wave ($e$-polarization), the RHS of the above
equation must be multiplied by $n_{1}^{2}/n_{2}^{2}$.

Substituting Eq.~(\ref{eq10}) into Eqs.~(\ref{eq7}) and (\ref{eq6}),
we can determine the functions
$\Phi_{\mathbf{m},\boldsymbol{\sigma}}\left( \omega_{s}\right)$,
which in turn determine the quantum state set forth in
Eq.~(\ref{eq5}). We consider a two-mode waveguide for which the
fundamental mode ($m_{q}=0$) is even and the next mode  ($m_{q}=1$)
is odd.  If the down-converted photons have different spatial
parity, the pump mode must be odd so that the spatial overlap
integral in Eq. (\ref{eq6}), which determines
$A_{\mathbf{m},\boldsymbol{\sigma}}\left( \omega_{s}\right)$ and
thence $\Phi_{\mathbf{m},\boldsymbol{\sigma}}\left(
\omega_{s}\right)$, does not vanish.

\begin{figure}
\centering
\includegraphics[width=3.4in,totalheight =4 in]{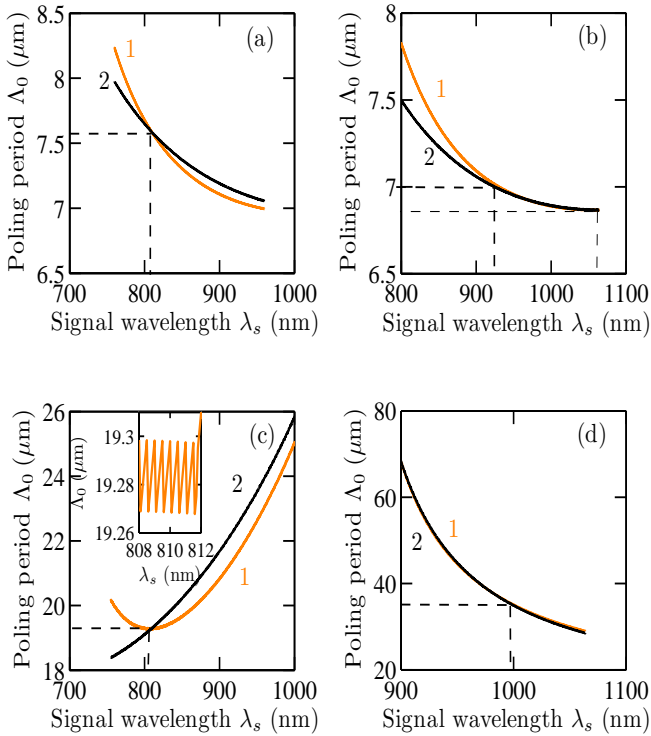}
\caption{(Color online). Values of the poling period $\Lambda_{0}$
required to satisfy quasi-phase matching as a function of the signal
wavelength $\lambda_{s}$, for four values of the waveguide thickness
$h$. Calculations are presented for a KTP planar waveguide with
$\Delta =0.05$ pumped by a laser whose wavelength is $\lambda_{p} =
532$ nm and whose modal structure coincides with $m_{p}=1$. The core
refractive indexes are determined using the Sellmeier equations for
KTP. The orange (black) curves represent condition 1 (2), namely
that the signal (idler) mode number $ m_{s}\:\left(m_{i} \right)$ is
0 and that the idler (signal) mode number $ m_{i}\:\left(m_{s}
\right)$ is 1. Dashed lines indicate intersection points or ranges.
(a) Type-0 $(e,e,e)$ with waveguide thickness $h$ = 1.625 $\mu$m.
(b) Type-0 $(e,e,e)$ with $h$ = 1.49 $\mu$m. (c) Type-II $(o,e,o)$
with $h$ = 1.617 $\mu$m. The inset shows the behavior of the poling
period for condition 1 near the intersection point. (d) Type-II
$(e,o,o)$ with $h$ = 1.56 $\mu$m.}\label{polingperiod}
\end{figure}

We seek to generate an entangled state of the form
\begin{equation}
\begin{array}{cl}
\vert \Psi\rangle\thicksim \displaystyle \int \!
\mathrm{d}\omega_{s}\!& \left[\Phi_{0,1,\boldsymbol{\sigma}}\left(
\omega_{s}\right)
\:\vert \omega_{s},0,\sigma_{s}\rangle\vert\omega_{i},1,\sigma_{i} \rangle\right. \\
 &\left.+\:\Phi_{1,0,\boldsymbol{\sigma}}\left( \omega_{s}\right)
 \:\vert\omega_{s},1, \sigma_{s}\rangle\vert\omega_{i},0,\sigma_{i}\rangle\right],
\end{array}
\end{equation}
i.e., if the signal is in the even mode, the idler must be in the
odd mode, and vice versa. This requires that the phase-matching
condition be satisfied for each of these two possibilities. For a
fixed geometry and material parameters, we find a single poling
period $\Lambda_0$ of the nonlinear coefficient at which these
conditions, $\Delta\beta_{0,1,\boldsymbol{\sigma}}\left(
\omega_{s}\right)  = 2\pi/\Lambda_0$ and
$\Delta\beta_{1,0,\boldsymbol{\sigma}}\left(  \omega_{s}\right)  =
2\pi/\Lambda_0$, are simultaneously met. We achieve this by plotting
the poling period $\Lambda_0$ as a function of the signal frequency
$\omega_s$, or as a function of the signal wavelength $\lambda_s$,
for each of these two conditions, and search for intersections. The
example shown in Fig.~\ref{polingperiod} reveals that this may occur
at a single frequency or, when the two curves are tangential, over a
broad spectral band. Once the poling period is selected and fixed,
we determine if the spectral functions
$\Phi_{0,1,\boldsymbol{\sigma}}\left( \omega_{s}\right)$ and
$\Phi_{1,0,\boldsymbol{\sigma}}\left( \omega_{s}\right)$ overlap,
i.e., have some common spectral band. If they do, then the
two-photon state exhibits entanglement.

\textbf{Example.} A laser source with wavelength $ \lambda_{p} =
532$ nm and in mode $m_{p}=1$, is used to pump a 1-D periodically
poled KTP waveguide with fractional refractive index change
$\Delta\approx1-n_{2}/n_{1} = 0.05$. The core refractive indexes
along the $y$ and $z$ axes are given by the KTP Sellmeier equations
\cite{KTPsellmeier}
\begin{equation}
\begin{array}{l}
n_{1,(y)}^{2}\left(\lambda \right) =3.45018+\dfrac{0.04341}{\lambda^{2}-0.04597}+
\dfrac{16.98825}{\lambda^{2}-39.43799},\vspace{2.5 mm}\\
n_{1,(z)}^{2}\left(\lambda \right) =4.59423+\dfrac{0.06206}{\lambda^{2}-0.04763}+
\dfrac{110.80672}{\lambda^{2}-86.12171},
\end{array}
\label{eq12}
\end{equation}
where $ \lambda $ is the wavelength expressed in $\mu$m.

The curves in Fig.~\ref{polingperiod} represent uniform
poling-period values satisfying the quasi-phase-matching conditions
versus the signal wavelength $\lambda_{s}$, for four different
values of the waveguide thickness $h$. The orange (black) curves
represent condition 1 (2), namely that the signal (idler) mode
number is 0 and the idler (signal) mode number is 1. In panels (a)
and (b), the interactions are Type-0 $(e,e,e)$, with $\left|d_{\rm
eff}\right|= \left|d_{33}\right|= 16.9$ pm/V
\cite{Handbooknonlinear}. The interactions in panels (c) and (d) are
Type-II $(o,e,o)$ and Type-II $(e,o,o)$, respectively, with
$\left|d_{\rm eff}\right|=\left|d_{24}\right|= 3.64$ pm/V. The
notation $(\cdot\,,\cdot\,,\cdot)$ indicates, in consecutive order,
the polarization of the down-converted photon whose frequency lies
above the degenerate frequency, the down-converted photon whose
frequency lies below the degenerate frequency, and the pump photon.
\begin{figure}
\centering
\includegraphics[width=3.4in,totalheight =4 in]{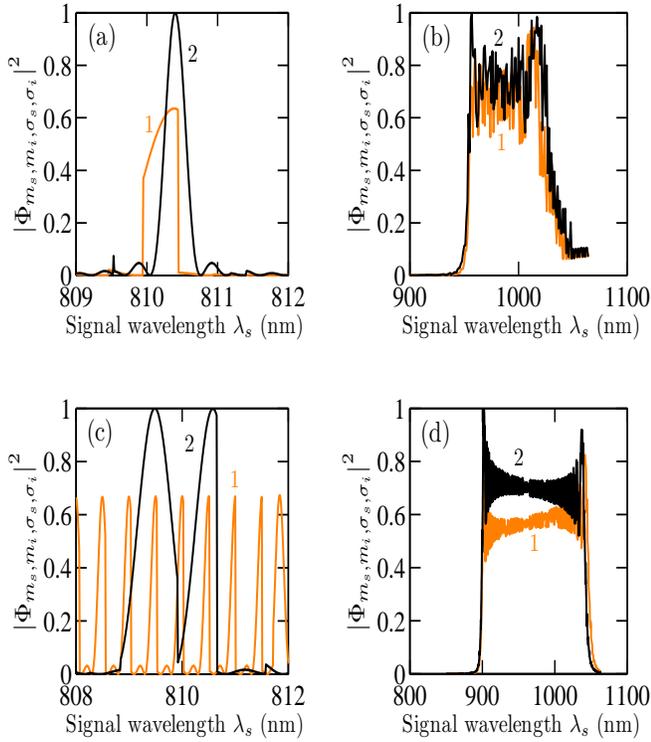}
\caption{(Color online). Normalized output spectra
$\left|\Phi_{\mathbf{m},\boldsymbol{\sigma}}\left(\omega_{s}\right)\right|^{2}
=\left|\Phi_{m_{s},m_{i},\sigma_{s},\sigma_{i}}\left(\omega_{s}\right)\right|^{2}$
as a function of the signal wavelength $\lambda_{s}$, for the four
entanglement conditions shown in Fig.~\ref{polingperiod}. Panels (a)
and (b) display modal/spectral entanglement regimes whereas panels
(c) and (d) display modal/spectral entanglement and
modal/polarization entanglement regimes. The orange (black) curves
represent condition 1 (2), namely that the signal (idler) mode
number $m_{s}\:\left(m_{i} \right)$ is 0 and that the idler (signal)
mode number $m_{i}\:\left(m_{s} \right)$ is 1. The waveguide length
$L$ is 25 mm. (a) Type-0 $(e,e,e)$, uniform poling with
$\Lambda_{0}$ = 7.5816 $\mu$m. (b) Type-0 $(e,e,e)$, linearly
chirped poling with $\Lambda_{0}$ = 6.87 $\mu$m and $\Lambda_{L}$ =
6.95 $\mu$m. (c) Type-II $(o,e,o)$, uniform poling with
$\Lambda_{0}$ = 19.298 $\mu$m. The multiple resonances that appear
in condition 1 result from the rippled behavior of the poling period
near the intersection point, as illustrated in the inset of
Fig.~\ref{polingperiod}(c). (d) Type-II $(e,o,o)$, linearly chirped
poling with $\Lambda_{0}$ = 30 $\mu$m and $\Lambda_{L}$ = 70
$\mu$m.}\label{spectra}
\end{figure}

These plots reveal that the two curves can intersect at a single
frequency, as in panels (a) and (c) or, alternatively, they can
coincide over a broad range of frequencies, as in panels (b) and
(d). Modal entanglement can be obtained at any intersection point
since the two quasi-phase-matching conditions are simultaneously
satisfied for the unique poling period determined by the point of
intersection. In panels (a) and (c), $h$ has been chosen such that
the intersection point between the curves occurs near $\lambda_{s}=
810$ nm so that the down-converted photon with the longer wavelength
lies in the telecommunications window near 1550 nm. In panels (b)
and (d), $h$ has been chosen such that the two curves are tangential
over a broad range of wavelengths.

We now proceed to study the spectral characteristics for the four
structures examined in Fig.~\ref{polingperiod}. The fixed poling
periods (spatial frequencies) established for panels (a) and (c) in
Fig.~\ref{polingperiod} are used for calculating the spectra
presented in panels (a) and (c) in Fig.~\ref{spectra}. Similarly,
the range of spatial frequencies where the curves overlap in
Figs.~\ref{polingperiod}(b) and (d) establish the linearly chirped
spatial frequencies used for the calculations displayed in
Figs.~\ref{spectra}(b) and (d). In all cases, we have chosen a
structure of length $L= 25$ mm.

The output spectra for the two conditions represented by the
quantities
$\left|\Phi_{0,1,\boldsymbol{\sigma}}\left(\omega_{s}\right)\right|^{2}$
and
$\left|\Phi_{1,0,\boldsymbol{\sigma}}\left(\omega_{s}\right)\right|^{2}$
are shown in Fig.~\ref{spectra}, normalized to the maximum of their
peak values. It suffices to plot these quantities solely for signal
frequencies above the degenerate frequency $\omega_{p}/2$ since
\begin{equation}
\Phi_{0,1,\sigma_{i},\sigma_{s}}\left(-\omega_{s}\right)=\Phi_{1,0,\sigma_{s},\sigma_{i}}\left(\omega_{s}\right).
\end{equation}
This relation holds by virtue of the definitions of
$A_{\mathbf{m},\boldsymbol{\sigma}}\left(\omega_{s}\right)$ and
$\Delta\beta_{\mathbf{m},\boldsymbol{\sigma}}\left(\omega_{s}\right)$.
Modal entanglement can be achieved over the overlapping regions of
the two spectra, as illustrated in Fig.~\ref{spectra}. Linearly
chirped poling leads to a broadband spectrum by virtue of the
continuum of its Fourier domain, so that multiple nonlinear
interactions can simultaneously be satisfied
\cite{carrasco04,nasr08}. There is, however, a tradeoff between
spectral breadth and photon flux density.

The astute reader will have noticed that we have excluded Type-I
interactions from our considerations. This is because the
down-converted photons have the same polarizations as those in
Type-0, but Type-I suffers from lower efficiency. The photon flux
from a Type-0 interaction is also expected to be greater than that
from a Type-II interaction since the former exploits the strongest
component of the second-order nonlinear tensor.

\section{Modal entanglement in 2-D circular waveguides (silica fibers)}
\begin{figure}
\centering
\includegraphics[width=2.8in,totalheight =1.9 in]{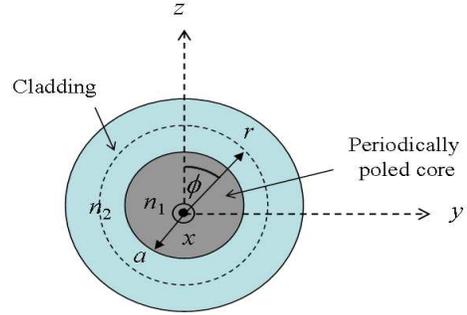}
\caption{Sketch of the cross-section of  a 2-D circular waveguide.
The waveguide is made up of a cylindrical rod of dielectric medium
with radius $\,a\,$ and uniform refractive index $n_{1}$ surrounded
by a medium of lower refractive index $n_{2}$. The inner and the
outer media are called the \emph{core} and the \emph{cladding},
respectively. The core is periodically poled in the $ x $
direction.} \label{fibersketch}
\end{figure}

Although silica is a centrosymmetric material, second-order
nonlinearities of $\sim$ 1 pm/V have been observed in poled silicate
fibers, see Fig. \ref{fibersketch}
\cite{glasspolingdeff,glasspoling,electricfieldglasspoling,planarglasspoling}.
The presence of a second-order nonlinearity in this centrosymmetric
material appears to stem from the intrinsic third-order nonlinear
susceptibility of the glass and the built-in electric field arising
from the displacement of charge species created during the poling
process. Although the effective nonlinear coefficient $\left|d_{\rm
eff}\right|$ in silica fiber is substantially smaller than that for
many second-order nonlinear materials (e.g., LiNbO$_{3}$ and KTP),
silica fibers are nonetheless of interest because they can be
fabricated in very long lengths. Second-order interactions in poled
glass fibers can be treated using the same techniques as those
developed for QPM structures.

In comparison with 1-D planar waveguides, which carry the index
$m_{q}$, each confined mode of a wave $q$ in a 2-D circular
waveguide is marked by an additional index, $l_{q}$. Based on the
geometry of the circular waveguide, cylindrical coordinates are a
natural choice for describing the transverse mode profile
\cite{salehteichbook}:
\begin{equation}
u_{l_{q}m_{q}}\left( \omega_{q},r,\phi\right) =
\left\lbrace
\begin{array}{cc}
 \mathrm{J}_{l_{q}} \! \left(k^{(T)}_{l_{q}m_{q}}r\right)\:\exp\left(j l_{q}\phi\right)&\quad\; r\leq a \vspace{2mm}\\
 \mathrm{K}_{l_{q}} \! \left(\gamma_{l_{q}m_{q}}r\right)\:\exp\left(j
l_{q}\phi\right)&\quad\; r>a.
\end{array}\right.
\label{eq13}
\end{equation}
Here $r$ and $\phi$ are the radial and azimuthal coordinates,
respectively; $l_{q}=0,\pm 1,\pm 2,\ldots$; $m_{q}=1,2,3,\ldots$;
$\mathrm{J}_{l_{q}}(\cdot)$ is the Bessel function of the first kind
of order $l_{q}$; $\mathrm{K}_{l_{q}}(\cdot)$ is the modified Bessel
function of the second kind of order $l_{q}$; $k_{l_{q}m_{q}}^{(T)}=
[\left(n_{1}\omega_{q}/c\right)^{2}-\beta_{l_{q}m_{q}}^{2}]^{1/2}$;
$\gamma_{l_{q}m_{q}}=
[\beta_{l_{q}m_{q}}^{2}-\left(n_{2}\omega_{q}/c\right)^{2}]^{1/2}$;
the superscript $(T)$ represents the transverse component of the
wavevector $k_{l_{q}m_{q}}$; $a$ is the radius of the fiber core;
and $n_{1}$ and $n_{2}$ are the refractive indexes of the core and
cladding, respectively (they are frequency-dependent).

The total number of modes $N_{q}$ is determined by the value of the
fiber parameter $V_{q}=(\omega_{q}\,a/c)(n_{1}^{2}-
n_{2}^{2})^{1/2}$. If, for example, $2.405<V_{q}<3.83$, then
$l_{q}=0,\pm 1$ and $m_{q}=1$, so that $N_{q}=3$. For each value of
$l_{q}$, the mode propagation constants $\beta_{l_{q}m_{q}}$ can be
determined by solving the fiber characteristic equation for
$k^{(T)}$ \cite{salehteichbook},
\begin{equation}
X
\dfrac{\mathrm{J}_{l_{q}\pm1}\left(X\right)}{\mathrm{J}_{l_{q}}\left(X\right)}
=\pm
Y\dfrac{\mathrm{K}_{l_{q}\pm1}\left(Y\right)}{\mathrm{K}_{l_{q}}\left(Y\right)},
\end{equation}
where $X=k^{(T)}a$ and $Y=(V_{q}^{2}-X^{2})^{1/2}$. The values of
$k^{(T)}$ that satisfy the characteristic equation are
$k^{(T)}_{l_{q}m_{q}}$. Using the definitions provided above, we can
then determine $\beta_{l_{q}m_{q}}$. The subscript
$\boldsymbol{\sigma}$ has been eliminated in this section since
glass fiber is an isotropic material and the characteristic equation
is polarization-independent.

In 2-D circular waveguides, the two-photon quantum state is
\begin{equation}
\begin{array}{c}
\vert \Psi\rangle\thicksim\displaystyle \int
\mathrm{d}\omega_{s}\:\sum_{\mathbf{lm}} \Phi_{\mathbf{lm}}\left(
\omega_{s}\right)\:\vert
\omega_{s},l_{s}m_{s}\rangle\vert\omega_{i},l_{i}m_{i}\rangle,
\end{array}
\label{eq14}
\end{equation}
where $ \mathbf{lm} =(l_{s}m_{s},l_{i}m_{i})  $, $ \Phi_{\mathbf{lm}} \left(\omega_{s}\right)$ can be
determined from Eq.~(\ref{eq7}) and the amplitude
$A_{\mathbf{lm}}\left(\omega_{s}\right)$ in cylindrical coordinates
is given by
\begin{equation}
\begin{array}{c}
A_{\mathbf{lm}}\left(\omega_{s}\right)=\displaystyle\int_{0}^{a}
\mathrm{d}r\int_{0}^{2\pi} \!\!\mathrm{d}\phi\; r
\displaystyle\prod_{q=p,s,i}
\!\!u_{l_{q}m_{q}}\left(\omega_{q},r,\phi\right).
\end{array}
\label{eq15}
\end{equation}
Substituting Eq.~(\ref{eq13}) into Eq.~(\ref{eq15}) and performing
the azimuthal integration leads to $\delta\left(l_{p}+l_{s}+l_{i}
\right)$, which can be viewed as a limitation on the transverse
profiles of the interacting modes (similar to that for planar
waveguides).

Consider nondegenerate SPDC in an optical fiber with
$2.405<V_{q}<3.83$. We seek to generate photon pairs with the
following entangled state:
\begin{equation}
\begin{array}{cl}
 \vert \Psi\rangle\thicksim \displaystyle\int \mathrm{d}\omega_{s}&
 \left[\Phi_{01,11}\left( \omega_{s}\right)\:\vert \omega_{s},01\rangle\vert\omega_{i},11\rangle\right. \vspace{1mm}  \\
 & \left. +\;\Phi_{11,01}\left( \omega_{s}\right)\:\vert
 \omega_{s},11\rangle\vert\omega_{i},01\rangle\right],
\end{array}
\end{equation}
signifying that if the signal is in the fundamental mode
$\left(l_{s}m_{s}=01\right)$, then the idler will be in the first
mode $\left(l_{i}m_{i}=11\right)$, and vice versa. To have
nonvanishing $A_{\mathbf{lm}}\left(\omega_{s}\right)$ or
$\Phi_{\mathbf{lm}} \left( \omega_{s}\right)$, the pump mode index
$l_{p}m_{p}$ must be $(-11)$.

\begin{figure}
\centering
\includegraphics[width=3.4in,totalheight =3.4 in]{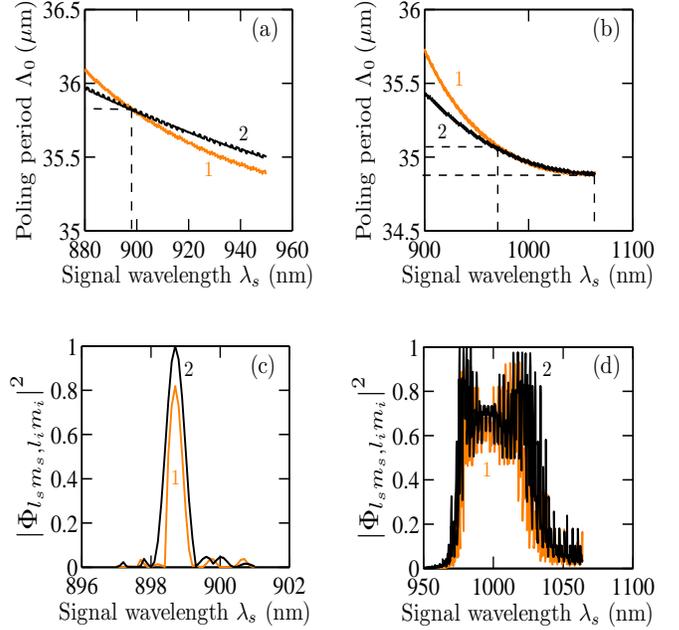}
\caption{(Color online).  A laser source with wavelength
$\lambda_{p} = 532$ nm and in mode $l_{p}m_{p}=-11$ is used to pump
a poled silica fiber with  $\Delta = 0.01$. (a) and (b) Poling
period versus the signal wavelength $\lambda_{s}$, with core radii
$a = 2.613$ and $a$ = 2.52 $\mu$m, respectively. The orange (black)
curve represents condition 1 (2), namely that the  signal (idler)
mode number $l_{s}m_{s}\,\left(l_{i}m_{i}\right)$ is 01 and the
idler (signal) mode number $l_{i}m_{i}\,\left(l_{s}m_{s} \right)$ is
11. Dashed lines show the intersection point or region. (c) and (d)
These panels display modal/spectral entanglement regimes. Normalized
output spectra
$\left|\Phi_{01,11}\left(\omega_{s}\right)\right|^{2}$ and
$\left|\Phi_{11,01}\left(\omega_{s}\right)\right|^{2}$ for the
structures displayed in (a) and (b), respectively, with $L = 20$ cm.
The poling period of the nonlinear coefficient is uniform in (c)
with $\Lambda_{0}$ = 35.813 $\mu$m, and is linearly chirped in (d)
with $\Lambda_{0}$ = 34.9 $\mu$m and $\Lambda_{L} $ = 35.05 $\mu$m.}
\label{fiberresults}
\end{figure}
\textbf{Example.} A pump source, with wavelength $\lambda_{p} = 532$
nm and in mode $l_{p}\,m_{p} = -11$, is incident on an optical glass
poled fiber with fractional refractive-index change $\Delta = 0.01$.
The core refractive index is computed using the Sellmeier equation
for fused silica \cite{salehteichbook}
\begin{equation}
\begin{array}{cl}
 n_{1}^{2}\left(\lambda \right) =& 1+\dfrac{0.6962\lambda^{2}}{\lambda^{2}-\left(0.06840\right)^{2}}
 +\dfrac{0.4079\lambda^{2}}{\lambda^{2}-\left(
 0.1162\right)^{2}}\\[5mm]
 & +\dfrac{0.8975\lambda^{2}}{\lambda^{2}-\left(9.8962\right)^{2}}\,,
\end{array}
\end{equation}
where $\lambda$ is the wavelength expressed in $\mu$m. Panels (a)
and (b) in Fig.~\ref{fiberresults} represent the dependence of the
uniform poling period on the signal wavelength $\lambda_{s}$
required to satisfy quasi-phase matching for the two conditions
under consideration, for two different values of the core radius
$a$. The curves can either intersect at a single frequency or they
can coincide over a broad range of frequencies [in analogy with the
results for the 1-D planar waveguide; see
Figs.~\ref{polingperiod}(a) and (b)]. Modal entanglement can be
achieved where the curves intersect. The core radius $a$ for panel
(a) has been chosen such that the intersection occurs at
$\lambda_{s} = 900$ nm, placing the down-converted photon with the
longer wavelength in the telecommunications window near 1300 nm. In
panel (b), $a$ has been selected such that the two curves are
tangential over a broad band of frequencies. Panels (c) and (d) in
Fig.~\ref{fiberresults} display the normalized output spectra
$\left|\Phi_{01,11}\left(\omega_{s}\right)\right|^{2}$ and
$\left|\Phi_{11,01}\left(\omega_{s}\right)\right|^{2}$ for fibers of
length $L  = 20$ cm. The poling periods of the fibers used in (c)
and (d) are determined from the results obtained in (a) and (b),
respectively. Modal entanglement can be achieved in wavelength
regions where the spectra overlap.

\section{Features of modal entanglement}
In the previous sections, we have demonstrated that it is possible
to obtain modal entanglement in 1-D planar and 2-D circular
waveguides. In this section, we discuss various features and
applications of modal entanglement.

\subparagraph{Combining the advantages of noncollinear and
collinear-degenerate interactions in bulk crystals.} A noncollinear
configuration is often preferred in bulk crystals because the photon
pairs generated by SPDC are easily separated. This process is
typically not efficient, however, since the fraction of the photons
that are entangled is seriously limited by the intersections between
the emission cones. A degenerate-collinear configuration, on the
other hand, creates impediments to separating the photon pairs but
offers high efficiency by virtue of the total overlap of the
emission cones. In a two-mode waveguide with modal entanglement, in
contrast, the photon pairs are efficiently generated in the two allowed
entangled modes and are thus also readily separated by their mode
numbers, i.e, their different transverse-field profiles. This can be
implemented by use of a branching waveguide \cite{modebranching}.

\subparagraph{Device flexibility and compatibility.}

Using uniform or linearly chirped poling, we are able to generate
either narrowband or broadband spectral/polarization entanglement,
as illustrated in Fig.~\ref{spectra}. By replacing the branching
waveguide at the device output with a diffraction grating, prism, or
polarizing beam splitter (for Type-II), we can obtain binary modal
entanglement. The structure can then be used as a source of binary
and continuum entanglement that is expected to be useful for
applications such as quantum imaging
\cite{abouraddy01,abouraddy02,nasr09}, quantum cryptography
\cite{jennewein00}, quantum teleportation \cite{bennett93}, and
quantum information \cite{nielsen00}. We note that this waveguide
configuration is compatible with integrated optics, which can
facilitate its incorporation into a practical system.

\subparagraph{Spectral entanglement} is typically measured via a
Hong-Ou-Mandel interferometer (HOM) \cite{HOM}. An experiment is conducted
by sweeping a temporal delay $\tau$ inserted between the
down-converted photons while measuring the coincidence rate of
photon counts at a pair of detectors placed at the two output ports
of the interferometer. The coincidence rate $R(\tau)$ is given by
\begin{equation}
R\left(\tau \right) =\textstyle\frac{1}{2}\left\lbrace
R_{0}-\mathrm{Re}\left[ R_{1}\left(\tau \right)\right] \right\rbrace
\end{equation}
with
\begin{equation}
\begin{array}{rl}
R_{0} =\displaystyle\int\mathrm{d}\omega_{s} & \left\vert\Phi_{\mathbf{m},\boldsymbol{\sigma}}\left( \omega_{s}\right) \right\vert^{2} , \\
R_{1}\left(\tau \right)= \displaystyle\int\mathrm{d}\omega_{s}& \left\lbrace \Phi^{}_{\mathbf{m},\boldsymbol{\sigma}}\left( \omega_{s}\right)\Phi_{\mathbf{m},\boldsymbol{\sigma}}^{*}\left( -\omega_{s}\right)\right.  \\
& \times\left. \exp\left[j\left( \omega_{p}- 2\omega_{s}\right)\tau \right]\right\rbrace ,
\end{array}
\end{equation}
where Re denotes the real part and the superscript $^*$ represents
the complex conjugate. The integration covers the entire spectrum,
so that either
$\Phi_{m_{0},m_{1},\boldsymbol{\sigma}}\left(\omega_{s}\right)$ or
$\Phi_{m_{1},m_{0},\boldsymbol{\sigma}}\left(\omega_{s}\right)$ can
be used to evaluate the integral, where $m_{0}$ and $m_{1}$ are the
two lowest modes of a waveguide. Again, changing the spatial mode
number is equivalent to changing the sign of the frequency, i.e.,
$\Phi_{m_{0},m_{1},\boldsymbol{\sigma}}\left( -\omega_{s}\right)
=\Phi_{m_{1},m_{0},\boldsymbol{\sigma}}\left( \omega_{s}\right)$.

\begin{figure}
\centering
\includegraphics[width=3.4in,totalheight =3.4 in]{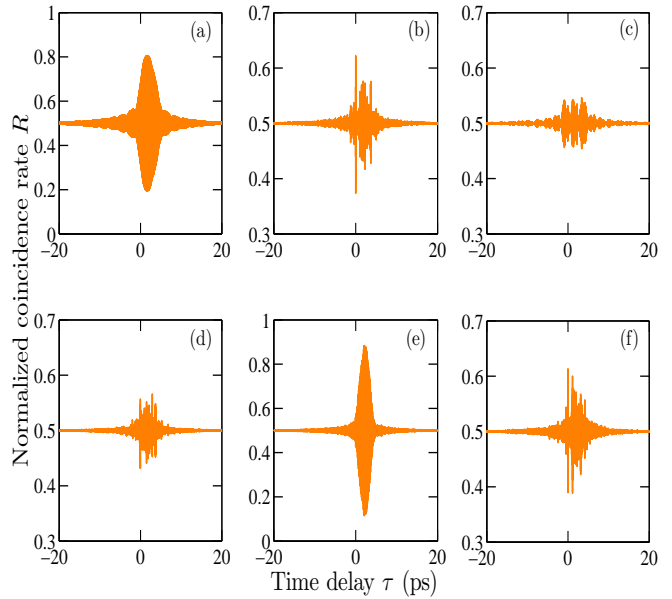}
\caption{(Color online). Dependence of the normalized coincidence
rate $R$ on the time delay $\tau$ between the two photons calculated
for a HOM interferometer. (a) 1-D planar waveguide with uniform
poling and Type-0 interaction. (b) 1-D planar waveguide with
linearly chirped poling and Type-0 interaction. (c) 1-D planar
waveguide  with uniform poling and Type-II interaction. (d) 1-D
planar waveguide with linearly chirped poling and Type-II
interaction. (e) 2-D circular waveguide  with uniform poling and
Type-0 interaction. (f) 2-D circular waveguide with linearly chirped
poling and Type-0 interaction. The structure and operational
parameters for panels (a), (b), (c), (d), (e), and (f) are the same
as those used for Fig.~\ref{spectra}(a), (b), (c), (d), and Fig.
\ref{fiberresults}(c) and (d), respectively.} \label{Coincidence}
\end{figure}
The graphs presented in Fig.~\ref{Coincidence} are the normalized
coincidence rate expected for an HOM interferometer for the planar
and circular waveguide structures described in Secs.~III and IV,
respectively. Since the signal and idler are nondegenerate, the
familiar dip in the coincidence interferogram is modulated by an
oscillatory function. When using the mode numbers for identifying
the photons, two different phase-matching conditions are required to
obtain spectral or polarization entanglement in waveguides, rather
than just a single one for bulk crystals. This distinction arises
because of the dependence of the propagation constant on the mode
number in a waveguide; in bulk crystals the magnitude of the
wavevector is independent of the direction of propagation. Hence,
$\Delta\widetilde{\beta}_{m_{0},m_{1},\boldsymbol{\sigma}}\left(\omega_{s}\right)$
is not exactly equal to
$\Delta\widetilde{\beta}_{m_{0},m_{1},\boldsymbol{\sigma}}\left(-\omega_{s}\right)$,
so that
$\Phi^{}_{\mathbf{m},\boldsymbol{\sigma}}\left(\omega_{s}\right)$
and
$\Phi_{\mathbf{m},\boldsymbol{\sigma}}^{*}\left(-\omega_{s}\right)$
do not perfectly overlap in the complex domain. This has several
consequences: $i)$ The interference pattern shifts about $\tau=0$
since
$\Phi_{\mathbf{m},\boldsymbol{\sigma}}\left(\omega_{s}\right)\Phi_{\mathbf{m},\boldsymbol{\sigma}}^{*}\left(-\omega_{s}\right)$
is a not purely real. This effect is illustrated in
Fig.~\ref{Coincidence}(a) and (e). $ii)$ The visibility, which is
defined as $\widetilde{V}\left(\tau\right) =\left(R_{\rm max}-R_{\rm
min}\right) /\left( R_{\rm max}+R_{\rm min} \right)$, where $R_{\rm
max}$ and $ R_{\rm min}$ are the maximum and minimum values of the
$R\left(\tau \right)$, respectively, is reduced since $R_{0}$ is not
precisely equal to $R_{\rm min}$. This is clearly
observable in the Type-II curves portrayed in Fig.
\ref{Coincidence}(c) and (d). $iii)$ The interference pattern is
ragged, as depicted in Fig.~\ref{Coincidence}(b) and (f). However,
reducing the device length will result in a cleaner interference
pattern, since the imaginary part of
$\Phi_{\mathbf{m},\boldsymbol{\sigma}}\left(\omega_{s}\right)\Phi_{\mathbf{m},\boldsymbol{\sigma}}^{*}\left(-\omega_{s}\right)$
is proportional to the device length for both uniform and linearly
chirped poling.

\begin{figure}
\centering
\includegraphics[width=3.4in,totalheight =1.9 in]{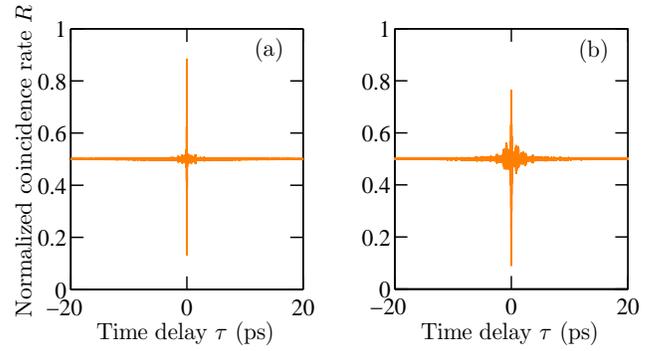}
\caption{(Color online). Dependence of the normalized coincidence
rate $R$ on the time delay $\tau$ between the two photons for a HOM
interferometer. (a) 1-D planar waveguide with linearly chirped
poling and Type-II interaction. (b) 2-D circular waveguide with
linearly chirped poling and Type-0 interaction. The structure and
operational parameters for panel (a) are the same as those used for
Figs.~\ref{spectra}(d) and \ref{Coincidence}(d) except that $L$ = 2
mm instead of 25 mm. The structure and the operational parameters
for panel (b) are the same as those used for
Figs.~\ref{fiberresults}(d) and \ref{Coincidence}(f) except that $L$
= 20 mm instead of 20 cm.} \label{shorter}
\end{figure}

Using Eqs.~(\ref{eq9}) and (\ref{eq8}), we define the
\textit{entanglement length} $L_{e}$ as the waveguide length for
which the phase of
$\Phi_{\mathbf{m},\boldsymbol{\sigma}}\left(\omega_{s}\right)\Phi_{\mathbf{m},\boldsymbol{\sigma}}^{*}\left(-\omega_{s}\right)$
is equal to $\pi$. For uniform poling, we obtain
\begin{equation}
L_{e}=\left\vert\frac{2\pi}{\Delta\widetilde{\beta}_{m_{0},m_{1},\boldsymbol{\sigma}}\left(\omega_{s}\right)-\Delta\widetilde{\beta}_{m_{0},m_{1},\boldsymbol{\sigma}}\left(
-\omega_{s}\right)}\right\vert,
\end{equation}
whereas for linearly chirped poling, we have
\begin{equation}
L_{e}=\left\vert\frac{4\pi^{2}\left(\Lambda_{0}^{-1}-\Lambda_{L}^{-1}
\right)}{\Delta\widetilde{\beta}^{2}_{m_{0},m_{1},\boldsymbol{\sigma}}\left(
\omega_{s}\right)-\Delta\widetilde{\beta}^{2}_{m_{0},m_{1},\boldsymbol{\sigma}}\left(
-\omega_{s}\right)}  \right\vert.
\end{equation}
To explicitly illustrate that the character of the interference
pattern does depend on the device length, we present plots of the
coincidence rate for shorter structures in Fig.~\ref{shorter}. The
visibility of the interference pattern increases and its
``noisiness'' decreases.

\begin{figure}
\centering
\includegraphics[width=3.4in,totalheight =1.8 in]{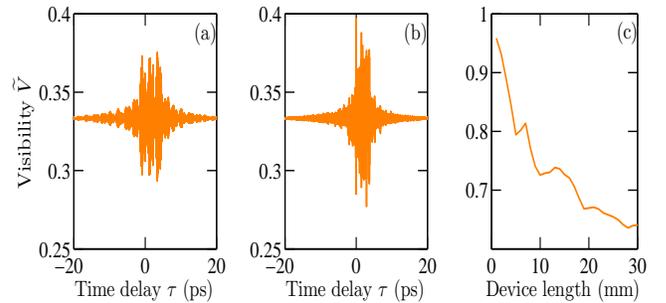}
\caption{(Color online). (a) and (b) Dependence of the visibility $
\widetilde{V}$ of the interference pattern on the time delay $ \tau
$ between the down-converted photons in Type-II interactions for 1-D
planar waveguides with uniform and linearly chirped poling,
respectively. The structure and operational parameters for panels
(a) and (b) are the same as those used for Fig.~\ref{spectra}(c) and
(d), respectively. (c) Dependence of the visibility $ \widetilde{V}$
on the device length $L$ for a 1-D planar waveguide with uniform
poling in a Type-II interaction. The parameters are the same as
those used for Fig.~\ref{polingperiod}(d), with a uniform poling
period $\Lambda_{0}$ = 67.7768 $\mu$m} \label{visibility}
\end{figure}

\subparagraph{Polarization entanglement} is measured by splitting a
pair of photons using a nonpolarizing beam splitter and then sending
them to two Glan-Thompson analyzers \cite{shih88}. After passage
through analyzers oriented at angles $\theta_{1}$ and $\theta_{2}$,
the coincidence rate turns out to be
\begin{equation}
R\left(\theta_{1}, \tau\right) =\textstyle\frac{1}{4}
\left(R_{0}-\textstyle\frac{1}{2}\sin^{2}\left(2\theta_1\right)
\left\lbrace R_{0}+\mathrm{Re}\left[R_{1}(\tau)\right]
\right\rbrace\right),
\end{equation}
where $\theta_{1} + \theta_{2}=90^\circ$; the angles are measured
with respect to the \emph{o}-polarization wave. The visibility of
the interference pattern is given by
$\widetilde{V}\left(\tau\right)= \left\lbrace
R_{0}+\mathrm{Re}\left[ R_{1}\left(\tau \right)\right]\right\rbrace
/\left\lbrace 3R_{0}-\mathrm{Re}\left[ R_{1}\left(\tau
\right)\right] \right\rbrace$. The dependence of the visibility on
temporal delay is displayed in Figs.~\ref{visibility}(a) and (b) for
structures that support Type-II interactions. The effect of the
waveguide length on the visbility is depicted in
Fig.~\ref{visibility}(c), which displays calculations carried out
for a 1-D KTP planar waveguide pumped using an \emph{o}-polarized
wave to generate Type-II $(e,o,o)$ down-conversion. The nonlinear
coefficient is taken to be uniformly poled with a period
$\Lambda_{0}$ = 67.7768 $\mu$m so that the wavelength of one of the
down-converted photons lies in one of the optical telecommunications
windows. The other parameters are the same as those used for
Fig.~\ref{polingperiod}(d). The visibility is seen to degrade with
increasing waveguide length. Hence, there is a tradeoff between long
waveguide length to generate a high flux of photon pairs and short
waveguide length to generate a smooth interference pattern with high
visibility.

\subparagraph{Modal entanglement} can be considered as an
alternative to polarization entanglement. Whereas polarization
entanglement is restricted to Type-II interactions, modal
entanglement has the merit that it can be generated using any type
of nonlinear interaction, including Type-0. Indeed, the
down-conversion generation rate in Type-0 is about one order of
magnitude higher than that in Type-II, since Type-0 utilizes the
strongest component of the second-order nonlinear tensor. Hence,
SPDC generated via Type-0 modal entanglement in a waveguide can
serve as an efficient source of binary entangled photons.

\subparagraph{An increase in the overall signal-to-noise ratio in an
infrared biphoton system} can be achieved by choosing the waveguide
dimensions such that one of the photons falls in the visible region
while the other lies in the infrared region. This technique can be
useful since the visible photon can be efficiently detected with a
Si photon-counting detector while the infrared photon is detected
with a less efficient InGaAs photon-counting detector. The overall
signal-to-noise ratio is increased as a consequence
\cite{abouraddy02,chatellus06}.

\subparagraph{Generation of a doubly entangled state via modal
entanglement.} Although Type-II is less efficient than Type-0, this
configuration can be exploited to generate a doubly entangled state
in frequency and polarization \cite{hyperentangled},
\begin{equation}
\begin{array}{c}
\vert \Psi\rangle\thicksim\left( \vert
\omega_{s},\omega_{i}\rangle+\vert
\omega_{i},\omega_{s}\rangle\otimes\vert
\sigma_{s},\sigma_{i}\rangle+\vert
\sigma_{i},\sigma_{s}\rangle\right).
\end{array}
\end{equation}
This state can be viewed as a combination of Type-II $ (o,e,o) $ and
Type-II $ (e,o,o) $. The photon pairs can be separated on the basis
of their mode number with the help of a branching waveguide. Using
Eq.~(\ref{eq51}), four different nonlinear processes are required to
generate the doubly entangled state:
\begin{equation}
\begin{array}{cl}
\vert \Psi\rangle\thicksim \displaystyle \int \mathrm{d}\omega_{s}&\left[\Phi_{0,1,o,e}\left( \omega_{s}\right)\:\vert \omega_{s},o\rangle_{0}\vert\omega_{i},e\rangle_{1}\right.\vspace{-1.5mm}\\
 &\left.  +\:\Phi_{1,0,o,e}\left( \omega_{s}\right)\:\vert\omega_{i},e\rangle_{0}\vert\omega_{s}, o\rangle_{1}\right. \\
 &\left.  +\:\Phi_{0,1,e,o}\left( \omega_{s}\right)\:\vert\omega_{s}, e\rangle_{0}\vert\omega_{i},o\rangle_{1}\right. \\
 &\left.  +\:\Phi_{1,0,e,o}\left( \omega_{s}\right)\:\vert\omega_{i},o\rangle_{0}\vert\omega_{s}, e\rangle_{1}
 \right].
\end{array}
\end{equation}
Although it is difficult to satisfy the phase-matching condition for
all of these processes using uniform poling, this can be achieved by
using linearly chirped poling. The nonlinear coefficient can also be
poled in an aperiodic sequence to obtain a narrowband doubly
entangled state \cite{norton04}.
\begin{figure}
\centering
\includegraphics[width=3.4 in,totalheight =1.8 in]{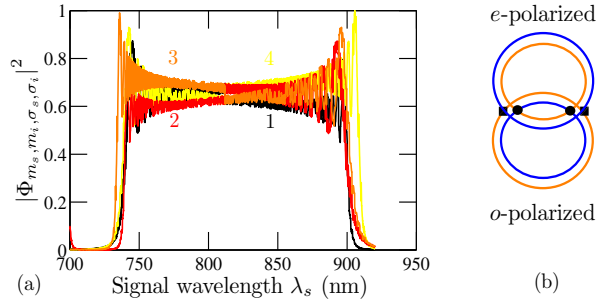}
\caption{(Color online). (a) Output spectra of SPDC from a KTP
planar waveguide with thickness $h$ = 1.1 $\mu$m, length $L$ = 25
mm, and $\Delta$ = 0.05, using an $o\,$-polarized source in mode
$m_{p}=1$ at $\lambda_{p} =$ 406 nm. The poling period of the
nonlinear coefficient is linearly chirped with $\Lambda_{0}$ = 6.5
$\mu$m and $\Lambda_{L}$ = 7.9 $\mu$m. The black (1) and red (2)
curves represent the output spectra associated with Type-II
$(o,e,o)$ modal entanglement, while the orange (3) and yellow (4)
curves represent the output spectra associated with Type-II
$(e,o,o)$ modal entanglement. The regions where the four spectra
overlap indicate modal/spectral/polarization entanglement. (b) The
regions of bulk Type-II $(o,e,o)$ and Type-II $(e,o,o)$ polarization
entanglement are marked by small black filled circles and squares,
respectively. The blue and orange rings represent photons with
frequencies above and below the degenerate frequency, respectively.}
\label{hyper}
\end{figure}

In Fig.~\ref{hyper}(a), we plot the normalized output spectra
$\left|\Phi_{m_{s},m_{i},\sigma_{s},\sigma_{i}}\right|^2$ when an  $
o\, $-polarized pump source in mode $ m_{p}=1 $ at $ \lambda_{p} =$
406 nm is incident on a 1-D KTP planar waveguide with thickness $ h
$ = 1.1 $\mu$m,  length $ L $ = 25 mm, and $ \Delta$ = 0.05. The
poling period of the nonlinear coefficient is linearly chirped with
$ \Lambda_{0} $ = 6.5 $\mu$m and $ \Lambda_{L} $ = 7.9 $\mu$m. As
illustrated in Fig.~\ref{hyper}, the output spectra associated with
Type-II $ (o,e,o) $ modal entanglement overlap with the output
spectra associated with Type-II $ (e,o,o) $ modal entanglement. The
region of overlap between the four spectra (except for the
degenerate case) represents the spectral band over which a doubly
entangled state can be obtained. It is worthy of note that this
state cannot be generated in bulk crystals since four different
directions are required, as shown in Fig.~\ref{hyper}(b). In
waveguides with modal entanglement, however, only two mode numbers
are required.

\section{Conclusion}
We have investigated nondegenerate spontaneous parametric
down-conversion (SPDC) in multimode 1-D planar and 2-D circular
waveguides. Using various types of interactions (Type 0 or II), the
generated photons can be entangled in the two lowest mode numbers of
the waveguide --- we refer to this as ``modal entanglement.'' The
inherent phase mismatch in the process is corrected by modulating
the nonlinear coefficient with poling periods (spatial frequencies)
that are either uniform or linearly chirped. Since the interaction
is collinear within the waveguides, modal entanglement can be used
in place of directional entanglement to distinguish between the
down-converted photons. Spectral or polarization entanglement can
therefore be generated with high efficiency and the entangled
photons can be readily separated. We find that there is a tradeoff
between long waveguide length to generate a high flux of photon
pairs and short waveguide length to generate a smooth interference
pattern with high visibility. In Type-0 interactions, frequency can
be used as a mode identifier rather than waveguide mode number,
offering the possibility of using modal entanglement in place of
polarization entanglement, which requires a Type-II interaction.
This is a salutary feature since a Type-0 interaction exploits the
strongest component of the second-order nonlinear tensor. The
technique can be implemented in any of the optical
telecommunications windows by controlling the waveguide dimensions.
Finally, we have demonstrated that modal entanglement in a Type-II
interaction can be used to generate a doubly entangled state in
frequency and polarization. Drawing on a Hilbert space of higher
dimensions can offer advantages in quantum-communication protocols
and quantum computing.


\begin{acknowledgments}
This work was supported by a U.S. Army Research Office (ARO)
Multidisciplinary University Research Initiative (MURI) Grant; and
by the Bernard M. Gordon Center for Subsurface Sensing and Imaging
Systems (CenSSIS), an NSF Engineering Research Center.
\end{acknowledgments}

\end{document}